# A mechanical engineer's approach to the measurement problem: Is a picture worth a thousand words?

Fred H. Thaheld

fthaheld@directcon.net

**Abstract**

The measurement problem is approached from a mechanical aspect, utilizing schematics to better assist in visualizing the boundary and the interplay between the quantum and classical worlds. This approach graphically illustrates what happens when a photon is absorbed by the retinal molecule, and reveals that even after collapse of the wave function, the *information* previously carried by the photon and passed on to the π* electron, is still at the microscopic level waiting to be amplified up to the macroscopic level by the rod cell. This analysis reveals that the measuring device or detector is of an unusual 2-stage nature, consisting of a microscopic entity or rhodopsin molecule which is not affected by the environment, contained within a macroscopic rod cell which serves as an amplifier. This approach may also allow one to better visualize the boundary conditions just *prior* to collapse and whether it is necessary to make any kind of adjustment to the Schrödinger equation to achieve a nonlinear collapse.

**Introduction**

A recent paper contained an interesting statement regarding Einstein's descriptions of Brownian motion, the photoelectric effect and general relativity from a mechanical aspect, showing his deep intuition about mechanics [1]. The author feels that Einstein is the greatest mechanical engineer of all times based upon these descriptions and, that the term "engineer" has no negative connotation, in that a physicist must be a good theorist and a good engineer.

It then occurred to me that maybe one could take a mechanical engineer's approach to the measurement problem, utilizing a simple schematic or drawing to assist us in determining which one of the many interpretations best fits this approach.

**The basic elements and their interaction**

Fig. 1 is a schematic of one of the ~$10^8$ rod cells that are in each human eye or retina, measuring 2 μ in dia x 100 μ in length [2-3]. To put this in perspective, the dia of a human hair varies from 80-120 μ [4]. There are ~$10^8$ rhodopsin molecules in each rod cell within the pigment containing discs. Rhodopsin consists of an opsin molecule containing ~ $4 \times 10^4$ nucleons and a dia of $4 \times 10^{-7}$ cm, with a retinal chromophore or light harvesting molecule embedded therein of 48 atoms or ~10 Å [5-6]. The outer segment contains the light transducing apparatus, and the inner segment contains the rod cell's nucleus and most of its biosynthetic machinery, while the synaptic terminal makes contact with the photoreceptor's target cells [2-3].

The structural chemical formula for 11-*cis*-retinal is shown in Fig. 2 [3]. When a photon of visible light is absorbed by the retinal molecule, a π orbital electron is excited to a π* orbital (a jump is made) and, within 200 fs, a conformational change of this molecule into all-*trans*-retinal has been completed. This means that it no longer fits in the opsin molecule, and that this change is *irreversible* [6]. The *trans* isomer separates from the opsin molecule and a series of changes begins. As this molecule changes its conformation, it initiates a cascade of biochemical reactions that result in the closing of $Na^+$ channels in the rod cell membrane [7]. Prior to this event $Na^+$ ions flow freely into the cell to compensate for the lower potential (more negative charge) which exists inside the rod cell. When the $Na^+$ channels are closed, a large potential difference builds up, with the inside of the cell becoming more negative as the outside of the cell becomes more positive [8]. This potential difference is passed along as a chemical signal of a nerve impulse. Thus it is that one photon activated rhodopsin molecule causes ~$10^6$ charges or $Na^+$ ions to fail to enter the rod cell, resulting in an amplified current about 1 pA in amplitude lasting ~200 ms [9]. This amounts to $2 \times 10^{-13}$ C, with $1.6 \times 10^{-19}$ C (which is the elementary unit of charge) per $Na^+$ ion [10]. This means that the *information* originally carried by the photon and passed on to the π electron and thence to the π* electron, has been amplified from the microscopic level and, that at some point in the amplification process, depending upon the number of $Na^+$ ions, crosses over into the macroscopic or classical world. This response results in between 2-3 action potentials in the optic nerve, all carrying this amplified information at the 1 pA level.

One does not have to know anything about chemistry to be able to grasp the importance of this change with regards to the measurement problem namely, that after the π-π* electron jump or collapse, the *information* transferred from the photon to the electron is *still at the quantum level*. Furthermore, that even after the first conformational change from *cis* to *trans*, we are *still at the quantum level* as the schematic graphically illustrates. The reason for this unusual sequence of counterintuitive events is because the measuring device is of a 2-stage nature, consisting of a microscopic $1^{st}$ stage or rhodopsin molecule, comprised of a retinal and opsin molecule, which is completely resistant to the environment or decoherence, contained within a macroscopic $2^{nd}$ stage or rod cell, where amplification of the collapsed or transferred *information* takes place. It is this amplification which brings the *information* out of the microscopic or quantum world to the macroscopic or classical world [6]. One does not have to invoke any need for the brain, mind or consciousness to somehow bring about a collapse of the wave function [11]. One can also now ascertain the exact boundary or the Heisenberg 'cut' between the microscopic and macroscopic worlds, which was heretofore considered unattainable [12-14].

**What about the Schrödinger equation and the linear to nonlinear transition?**

The next issue has to do with the very beginning of this process, just before the photon is absorbed by the retinal molecule, and addresses the difficult problem of whether any adjustment has to be made to the Schrödinger equation so that the transition from linear to nonlinear can be accomplished. Can the schematic in Fig. 2 assist us in this analysis?

First off, one has to ask the question as to just what else can a wave function be expected to do when it comes into contact with a retinal molecule, in that in the final analysis it must be a photon which is absorbed by this molecule? This must be a very simple and generic process, in that it takes place every second of every day among ~1.3 x $10^6$ other living entities, i.e. in the visual receptors of all three phyla which possess eyes: mollusks, arthropods and vertebrates, and has been doing so for millions of years [15-16]! This means that we all see essentially the same collapsed information trillions of times per second no matter where we are in the world if we are observing the same event. Are those two dynamical principles, the Schrödinger linear evolution and the von Neumann nonlinear projection or collapse postulate really incompatible? This paper has revealed already the existence of an identifiable borderline between these two different regimes and, that the transition from linear to nonlinear, deterministic to stochastic, reversible to irreversible, and quantum to classical, appears to be a very *natural* process which requires *no modification* to the Schrödinger equation.

Let us first take a look at von Neumann's two-process projection postulate and see if it might cast some light upon the issue at hand [17]. The 1st process states that light or the measured quantum system *S*, interacts with a macroscopic measuring apparatus *M* (the photoreceptor rhodopsin molecule-rod cell) for some physical quantity *Q*, with the interaction governed by the linear, deterministic Schrödinger equation [17-18]. For the time being we will say that *M* is macroscopic, although the retinal molecule embedded within it is microscopic, as has been previously pointed out. The 2$^{nd}$ process states that after this first stage of the measurement terminates, and one has a linear combination of products which are called entangled states, a 2$^{nd}$ nonlinear, indeterminate process takes place, the collapse of the wave packet.

To carry this analysis a little further, the idea of *naturalization* of the collapse (when taken in league with the *naturalization* of the Born rule), underlies the whole issue of quantum mechanics [19]. "One strategy which is being attempted with regard to this issue, is to focus upon the physical act of synchronization, which means that a certain branch of the wave function is always chosen because it is synchronous with the object upon which it impinges and collapses [20]. If the act of synchronization has already been completed, the descriptive scheme of synchronous nature could be vindicated as in the case of the Schrödinger equation. In contrast, if the action for synchronization on the spot comes to the surface, this must be the place being full of any sort of the wave function collapse and the like. The action for synchronization must be intrinsically nonlinear, while synchronous movement could be linear [20]. The other side of the coin is this: Once we are determined to accept the ubiquity of the linear Schrödinger scheme, we must also be prepared to accept on the part of nature, some sort of nonlinear acts for picking up something linear in the hindsight" [20].

In addition, this approach embraces what is known as the *internalist* stance proposed by Matsuno [21-22]. This means that the material act of distinguishing between before and after physical events, whatever they are, to be most fundamental, irreducible and even ubiquitous inside this empirical world. The linear approach, no matter how cherished by the majority, would remain secondary at best. Once one accepts this stance, nonlinearity intrinsic to the internal act of making distinctions, would turn out to be the rule rather than the exception, which will represent a new doctrine [19, 21-22].

**Conclusion**

1. It appears that the standard Copenhagen interpretation best fits this model vs all the other interpretations, whether one accepts the idea of a 2-stage measuring device with its microscopic and macroscopic segments, or considers it all as just a macroscopic device.

2. We can now identify a definite borderline between these two dynamical principles, whether it be linear to nonlinear, deterministic to stochastic, reversible to irreversible or quantum to classical.

3. That one can now accurately establish the place where the Heisenberg 'cut' takes place between the microscopic and macroscopic worlds. One has a choice to either insert this 'cut' at the retinal molecule or during the amplification process in the rod cell.

4. That a definite collapse of the wave function can be shown to take place when the $\pi$ electron jumps to a $\pi^*$ orbital electron.

5. That the linear to nonlinear transition of the Schrödinger equation is a *natural* process when a wave function comes into contact with a retinal molecule. There is no other choice for it but to make a transition to a photon, since a wave function cannot be absorbed by this molecule, only a photon of light.

6. That von Neumann's two-process projection postulate appears to best fit items 3 and 4 based on some slight adjustments.

7. It may well be that the solution to the measurement problem will be as simple, in retrospect, as were Einstein's explanations of the photoelectric effect, Brownian motion and general relativity. So it is that one arrives at an interdisciplinary answer to the measurement problem involving physics, chemistry, biology and philosophy.

**Acknowledgement**

This is dedicated to my father, who was one of the best mechanical engineers I ever knew from the standpoint of designing planes, diesel engines and oil tools. This exposure has enabled me to view biology and physics from a unique mechanical perspective.

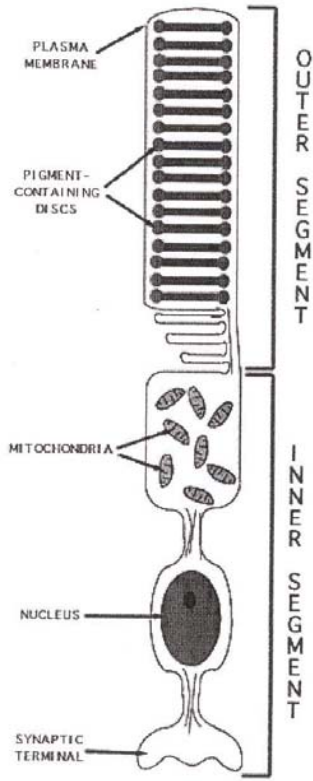

Fig. 1

Rod cell

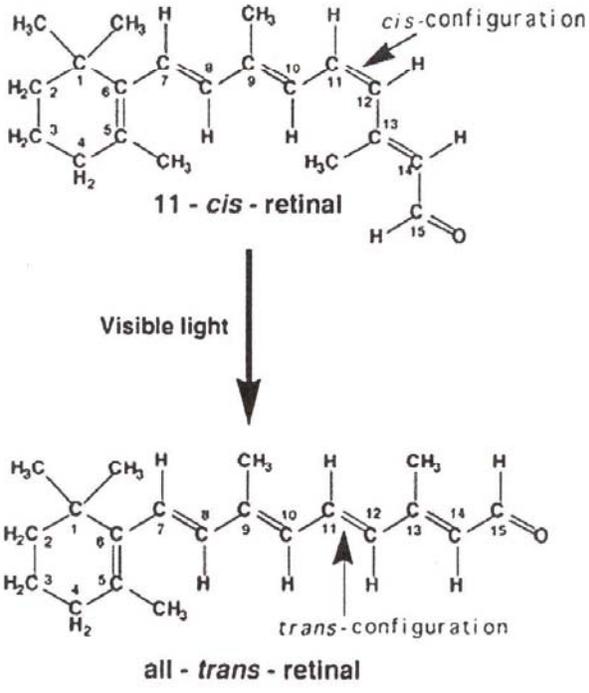

Fig. 2

Retinal molecule